\documentclass[acmsmall,screen,nonacm]{acmart}
\setcopyright{none}               

\usepackage{graphicx}
\usepackage{subcaption}
\usepackage{xcolor}
\usepackage{colortbl}
\usepackage{diagbox} 
\usepackage{wrapfig}
\usepackage{enumitem} 
\usepackage{tabularx}
\usepackage{xcolor,soul,framed} 
\usepackage{multirow,diagbox}
\colorlet{shadecolor}{yellow}
\usepackage{algorithm}
\usepackage{algorithmic}
\usepackage{amsmath}
\usepackage{float}

\usepackage[most]{tcolorbox}
\newtcolorbox{myresultbox}[1][]{
    colback=gray!20,
    colframe=black!75,
    fonttitle=\bfseries,
    title=Result,
    #1
}

\AtBeginDocument{%
  }

\setcopyright{acmcopyright}
\copyrightyear{2018}
\acmYear{2018}
\acmDOI{XXXXXXX.XXXXXXX}
\acmJournal{JACM}
\acmVolume{37}
\acmNumber{4}
\acmArticle{111}
\acmMonth{8}

\newcommand{\myautoref}[2]{\hyperref[#2]{#1~\ref*{#2}}}

\makeatletter
\renewcommand\paragraph{\@startsection{paragraph}{4}{0pt}%
  {-.2\baselineskip \@plus -2\p@ \@minus -.2\p@}%
  {-3.5\p@}%
  {\ACM@NRadjust{\bfseries\itshape\@adddotafter}}}
\makeatother

\ifx\formal\undefined
  
\else
   
\fi

\newcommand{\bb}{\begin{array}{llllllll}}
\newcommand{\ee}{\end{array}}

\begin{document}

\title{ VulAgent: A Hypothesis Validation-Based Multi-Agent System for Software Vulnerability Detection}

\author{Ziliang Wang}
\email{wangziliang@pku.edu.cn}
\author{Ge Li}
\authornote{Corresponding author.}
\email{lige@pku.edu.cn}
\affiliation{%
  \institution{Key Lab of High Confidence Software Technology, MoE, School
of Computer Science, Peking University}
  \city{Beijing}
  \country{China}
}

\author{Jia Li}
\email{jia_li@mail.tsinghua.edu.cn}
\affiliation{%
  \institution{College of AI, Tsinghua University}
  \city{Beijing}
  \country{China}
}

\author{Hao Zhu}
\email{zhuhao@stu.pku.edu.cn}

\author{Zhi Jin}
\email{zhijin@sei.pku.edu.cn}

\affiliation{%
  \institution{Key Lab of High Confidence Software Technology, MoE, School
of Computer Science, Peking University}
  \city{BeiJing}
  \country{China}
}

\renewcommand{\shortauthors}{Trovato et al.}
\renewcommand\footnotetextcopyrightpermission[1]{} 

\begin{abstract}
The application of language models to project-level vulnerability detection remains challenging, owing to the dual requirement of accurately localizing security-sensitive code and correctly correlating and reasoning over complex program context.
We present VulAgent, a multi-agent vulnerability detection framework based on hypothesis validation. Our design is inspired by how human auditors review code: when noticing a sensitive operation, they form a hypothesis about a possible vulnerability, consider potential trigger paths, and then verify the hypothesis against the surrounding context.
%
VulAgent implements a semantics-sensitive, multi-view detection pipeline: specialized agents, each aligned to a specific analysis perspective (e.g., memory, authorization), collaboratively surface and precisely localize sensitive code sites with higher coverage.
Building on this, VulAgent adopts a hypothesis-validation paradigm: for each vulnerability report, it builds hypothesis conditions and a trigger path, steering the LLM to target the relevant program context and defensive checks during verification, which reduces false positives.
On average across the two datasets, VulAgent improves overall accuracy by 6.6\%, increases the correct identification rate of vulnerable–fixed code pairs by up to 450\% (246\% on average), and reduces the false positive rate by about 36\% compared with state-of-the-art LLM-based baselines.

\end{abstract}





\maketitle

\section{Introduction}
Software vulnerabilities, caused by insecure coding practices, represent critical weaknesses that attackers can exploit to compromise software systems, leading to severe consequences such as unauthorized information disclosure~\cite{fu2022linevul} and cyber extortion~\cite{thapa2022transformer}.
The scale of this threat continues to rise: in the first quarter of 2022, the US National Vulnerability Database (NVD) disclosed 8,051 vulnerabilities, a 25\% increase over the previous year~\cite{cheng2022path}, while a Synopsys OSSRA study reported that 81\% of 2,409 analyzed codebases contained at least one known open-source vulnerability.
These statistics highlight the urgent importance of effective vulnerability detection mechanisms to strengthen software security and mitigate potential risks~\cite{fu2022linevul,thapa2022transformer,jang2014survey,johnson2011guide,cao2025vulpa}.

\begin{figure*}[h]
\centering
\includegraphics[width=0.68\textwidth]{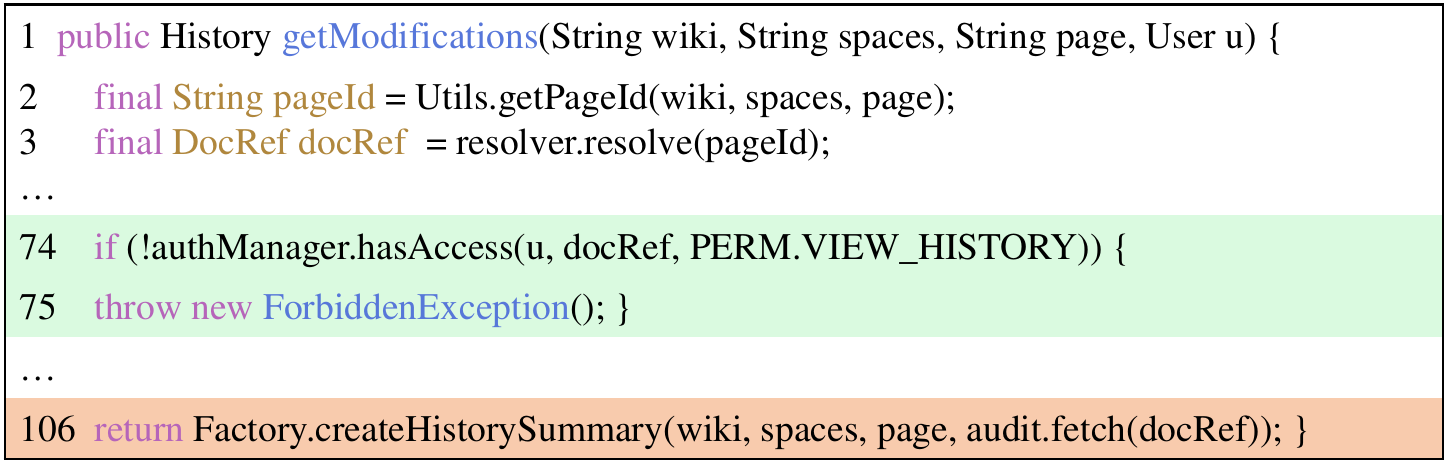}

\caption{A simplified code snippet for CWE-862 (Missing Authorization).
}
\label{fig:code}
\end{figure*}

\textbf{Problem Statement. } This paper is committed to addressing the challenge of how to use large language models to detect context-sensitive vulnerabilities with low false positive rates.

Figure~\ref{fig:code} shows a simplified example.
The function \texttt{getModifications(wiki, spaces, page, u)} builds a resource id at L2:
\texttt{pageId = Utils.getPageId(...)}.
It resolves this id to \texttt{docRef} at L3.
The code checks access at L74--L75 with
\texttt{authManager.hasAccess(u, docRef, PERM.VIEW\_HISTORY)}.
If the check fails, it returns or throws.
Several helper or log calls follow; we omit them (L76--L105).
At L106, the code invokes a sensitive operation:
\texttt{Factory.createHistorySummary(wiki, spaces, page, audit.fetch(docRef))}.
If the check at L74--L75 is missing or bypassed, an unauthenticated user can reach L106
and read hidden data. 
This is a typical \texttt{CWE-862} case.

From the above process, it can be seen that accurate vulnerability detection requires a detector to surface risk-bearing code—such as permission-related operations (such as L106)—and then use program context to associate these operations with their definitions and required authorization (such as L74-75).
When the detector fails to surface such code, true issues are missed and false negatives arise.
When it cannot reliably recognize the check or link it to the same operation in context, many false positives result.
Therefore, the core challenge is:
\emph{How can we accurately locate risky code and effectively use program context for vulnerability detection?}

\textbf{Prior Work. } 
%
Existing LLM-based detection methods struggle to mitigate the above problem.
Single-agent approaches based on chain-of-thought (CoT) were first explored for vulnerability detection, but the lack of explicit context modeling and false-positive (FP) filtering makes performance hard to accept~\cite{ding2025vulnerability}.
Li et al.~\cite{li2025everything} proposed concatenating program context with the detection prompt; however, merging detection with context verification increases reasoning depth and coupling, and FP rates remain high.
\emph{iAudit}~\cite{ma2024combining} assigns \emph{detector}, \emph{reasoner}, and \emph{critic} roles: the detector surfaces candidate issues (sensitive code discovery), the reasoner elaborates, and the critic filters;
%
%
\emph{MuCoLD}~\cite{mao2024multi} conducts a tester–developer discussion to refine the initial decision and reduce FPs.
\emph{VulTrial}~\cite{widyasari2025let} adopts a courtroom-style process with a third-party judge to arbitrate between detector and evaluator; 
Reflection- or discussion-based filters focus on generated text rather than reconstructing path-level evidence, which limits effective use of program context;
Overall, the current detection methods, which rely on single-agent detection and a reflection mechanism for generated content, are unable to meet the requirements of both aspects.

\textbf{This Work.} We have proposed two novel designs to address this challenge:

\begin{itemize}[leftmargin=*,topsep=2pt,itemsep=2pt]
\item \textbf{Multi-view, semantics-driven vulnerability detection.} A MetaAgent routes the code to specialized analyzers (e.g., memory safety, permission/authorization, file I/O, concurrency) to surface sensitive operations and candidate issues from multiple views, then an aggregator deduplicates and merges the reports. This decouples the detection task so each agent focuses on a single concern, lowers reasoning complexity, and increases coverage of sensitive statements, reducing missed detections.

\item \textbf{Hypothesis Validation filtering under program context.} For each candidate, a Path Planner constructs a vulnerability hypothesis consisting of the hypothesis conditions and the hypothesis path.
A Hypothesis-conditions Validation Agent then uses context retrieved from static tools (e.g., Joern) to perform targeted searches guided by the hypothesis conditions and to test the hypothesis’ plausibility, rejecting unreasonable cases as the first pruning stage. 
Reports with reasonable hypotheses are passed to a Final Validator, which, assuming the hypothesis holds, checks whether the hypothesis path actually exists and whether protections on that path (e.g., early returns, error handling, boundary checks) prevent exploitation; non-exploitable cases are removed and the final detection report is produced.
\end{itemize}

\textbf{Contribution. }
Based on the above process, we introduce a hypothesis validation mechanism.
This design enables large language models to purposefully leverage program context information, rather than relying on naïve concatenation, thereby constructing a more reliable vulnerability detection pipeline.
To evaluate its effectiveness, we conduct extensive experiments on two representative datasets: PrimeVul  and SVEN, demonstrating the superiority of our approach over state-of-the-art baselines.
In summary, the main contributions of this paper are as follows:
\begin{itemize}
\item [a)]We propose VulAgent, a multi-agent vulnerability detection framework that decouples the detection pipeline into coordinated stages: discovery, report aggregation, hypothesis construction, hypothesis validation. This modular design provides a systematic and principled approach to vulnerability detection, moving beyond monolithic strategies.

\item [b)] We introduce a hypothesis validation–based filtering mechanism, where vulnerability hypotheses are structured with explicit triggering conditions and execution paths. By validating these hypotheses against program context and defensive mechanisms, VulAgent enables large language models to leverage context in a targeted manner, thereby substantially reducing false positives while retaining high recall.

\item [c)]
VulAgent outperforms strong baselines across PrimeVul and SVEN, improving average accuracy by 6.6\%, increasing correct identification of vulnerable–fixed pairs by up to 450\% (246\% on average), and reducing the false-positive rate by approximately 36\%, demonstrating robust gains.

\end{itemize}

\section{Approach}
\begin{figure*}
\centering
\includegraphics[width=0.98\textwidth]{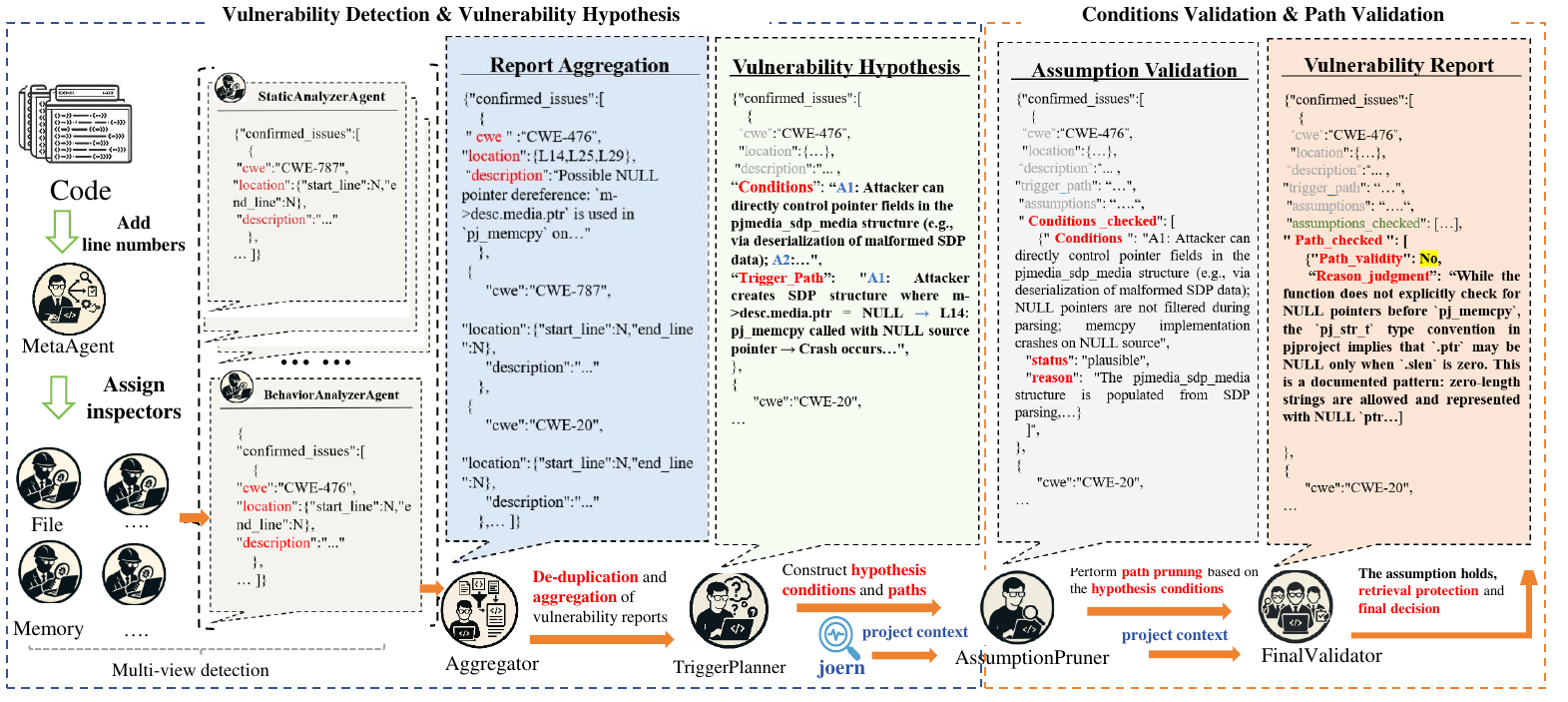}
\caption{The framework of VulAgent, which coordinates multiple specialized agents for vulnerability detection, aggregates their reports into vulnerability hypotheses, and then performs hypotheses validation with program context.
}
\label{fig:main}
\end{figure*}

In this section, we present the architecture of VulAgent, shown in Figure~\ref{fig:main}.
VulAgent decouples vulnerability detection into four stages: multi-view detection and integration, hypothesis construction, hypothesis validation, and protection checking.

%

\subsection{Phase I:  Multi-view Vulnerability Detection}

We denote the dataset as $D = { (c_i^{vul}, c_i^{fix}, y_i, m_i, w_i) }_{i=1}^N$, where $c_i^{vul}$ is a vulnerable code version, $c_i^{fix}$ is the corresponding fixed version, $y_i$ is the vulnerability label, $m_i$ denotes the project name, and $w_i$ represents the CWE category. Here, $N$ is the total number of code pairs. 
Each pair therefore contains both the vulnerable and patched implementations, together with their security label, project context, and standardized vulnerability type, which serve as the basis for our evaluation.

\textbf{Preprocessing Process}: We preprocess the input code by assigning line numbers to each line. 
For example, a line int foo() becomes L1: int foo(). The numbered format allows sensitive lines to be consistently referenced and propagated across agents during detection, validation, and reporting.

\begin{wrapfigure}{l}{0.48\textwidth}
  \centering
  \vspace{-0pt}
  \includegraphics[width=0.48\textwidth]{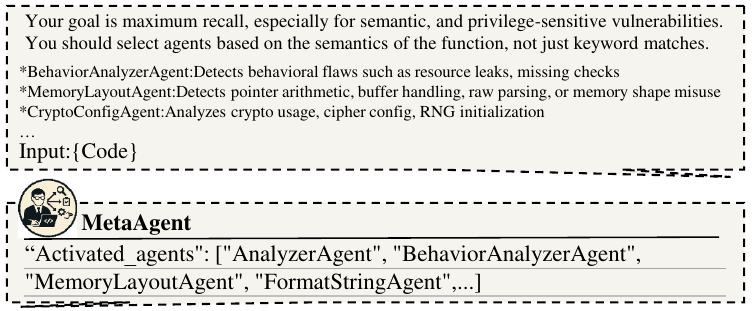}
  \label{fig:metaagent}
  \vspace{-20pt}
\end{wrapfigure}
\textbf{Assign Inspectors.}
The MetaAgent assigns specialized checking agents to the code under analysis.
Given a code unit $c_i$, it first parses structure and extracts semantic cues, including memory operations, file system calls, concurrency primitives, cryptographic APIs, and privilege-related logic.
Based on these cues, it activates a \emph{minimal-but-sufficient} set of agents for analysis.

To guarantee basic coverage, the MetaAgent always activates three complementary agents:
\emph{StaticAnalyzerAgent} (syntax-level pattern scanning for obvious red flags),
\emph{BehaviorAnalyzerAgent} (control-/data-flow reasoning to expose path and error-handling flaws),
and \emph{MemoryLayoutAgent} (pointer arithmetic and buffer boundaries).
These three form a safety net even when other specialized agents (e.g., crypto or auth) are not triggered.

Let $\mathcal{B}=\{\text{AnalyzerAgent},\text{BehaviorAnalyzerAgent},\text{MemoryLayoutAgent}\}$ be the baseline set, and let $\phi(c_i)$ denote the code being tested.
Let $R$ denote the LLM-based semantic dispatcher (MetaAgent) that maps code semantics (and optional context) to agent activations.
The activated set $A_i$ is
\begin{equation}
  A_i = \mathcal{B} \mathrel{\cup} \Bigl( \bigcup_{r\in R} r\!\bigl(\phi(c_i)\bigr) \Bigr),
  \qquad 0 < i \le N .
\end{equation}

\begin{wrapfigure}{l}{0.48\textwidth}
  \centering
  \vspace{-10pt}
  \includegraphics[width=0.48\textwidth]{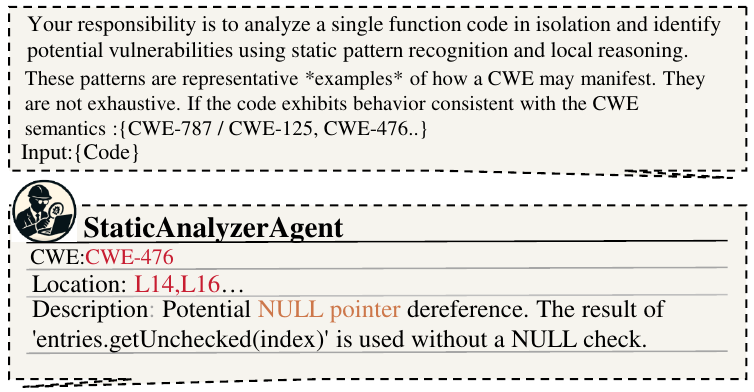}
  \label{fig:baseagent}
  \vspace{-20pt}
\end{wrapfigure}
For example,
\begin{equation}
  \hspace*{1.02\linewidth}
\begin{aligned}
  &\texttt{format\_string}\;\mapsto\;\text{FormatStringAgent}.
\end{aligned}
\end{equation}

\noindent\textbf{Specialized agents.}
We design a set of focused agents, each concentrating on a specific class of vulnerabilities, so the workload is modularized and multi-view collaboration reduces single-model bias:
\begin{enumerate}
  \item \textbf{StaticAnalyzerAgent}: baseline static scanning for suspicious patterns.
  \item \textbf{BehaviorAnalyzerAgent}: control/data-flow simulation for runtime risks.
  \item \textbf{MemoryLayoutAgent}: pointer arithmetic and buffer safety.
  \item \textbf{FormatStringAgent}: uncontrolled format string usage.
  \item \textbf{FilePermissionAgent}: unsafe file operations and permission changes.
  \item \textbf{AuthFlowAgent}: authentication and privilege logic.
  \item \textbf{CryptoConfigAgent}: weak or misconfigured cryptographic operations.
  \item \textbf{ConcurrencyAnalyzerAgent}: races and synchronization issues.
  \item \textbf{ErrorHandlingAgent}: missing error handling or resource cleanup.
  \item \textbf{CodeInjectionAgent}: dynamic code execution and command injection.
\end{enumerate}

All base agents share a uniform prompt template so that their outputs are consistent and easy to aggregate.
Each prompt contains: 
(i) \emph{Role \& Task} — what risk the agent targets (e.g., memory layout, format string, file permission, auth flow, crypto config, concurrency, error handling, code injection) for a \emph{single, line-numbered C/C++ function};
(ii) \emph{CWE focus \& trigger hints} — primary CWE IDs plus typical triggers; when applicable;
(iii) \emph{Output contract} — a STRICT JSON schema; 
and (iv) \emph{Inputs} — the numbered function body (\texttt{\{function\_code\}}) and optional context.
Based on this design principle, the basic agents can be quickly constructed and registered in MetaAgent to meet the requirements of different projects.

Once activated, each agent $a\in A_i$ analyzes $c_i$ from its angle and emits a textual report $t_i^a$ (finding, trigger, and evidence).
Because agents operate conservatively to maximize recall, the combined reports contain many false positives; therefore, the raw outputs serve as candidates that must be refined, pruned, and validated in subsequent stages (hypothesis construction and validation).

\begin{wrapfigure}{l}{0.48\textwidth}
  \centering
  \vspace{-8pt}
  \includegraphics[width=\linewidth]{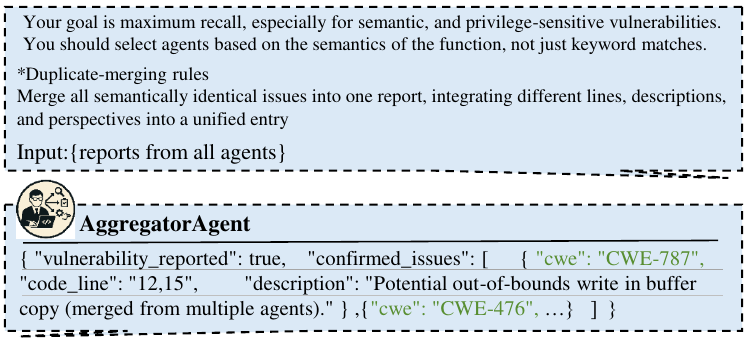}
  \label{fig:aggregator}
  \vspace{-10pt}
\end{wrapfigure}

\noindent\textbf{Report aggregation.}
Let
\begin{equation}
  \hspace*{1.05\linewidth}
  T_i = \{\, t_i^a \mid a \in A_i \,\}.
\end{equation}
The \emph{AggregatorAgent} merges overlapping findings from different agents and produces a de-duplicated set
\begin{equation}
  \hspace*{1.05\linewidth}
  \widetilde{T}_i \;=\; \mathrm{Agg}(T_i).
\end{equation}
Aggregation follows simple rules consistent with our system:
\begin{itemize}
  \item \textit{Group by CWE and span.} Items that report the same CWE and refer to the same or overlapping line range are grouped together; others remain separate.
  \item \textit{Merge within a group.} Keep one concise description (choose the most representative phrasing), union the evidence snippets and line references, and record the list of \texttt{source\_agents}.
  \item \textit{Preserve essentials.} Do not drop items that differ in CWE or non-overlapping spans; they are carried forward as individual entries.
\end{itemize}

\noindent The aggregated set $\widetilde{T}_i$ is then passed to the \emph{TriggerPlannerAgent} in Phase~II to construct structured vulnerability hypotheses (conditions and trigger paths).

\subsection{Phase II: Vulnerability Hypothesis Construction}

The goal of this phase is to transform aggregated vulnerability reports into structured vulnerability hypotheses. 
%
A vulnerability hypothesis $h_i$ is composed of two parts:

\begin{itemize}
  \item \textbf{Hypothesis Conditions}: these specify the preconditions under which a vulnerability might exist (e.g.,``the buffer length may exceed its allocated size'', ``the input pointer may be NULL''). 
 Such hypotheses  are derived from the static characteristics of the code and the descriptions provided by the detection agents.

  \item \textbf{Hypothesis Path}: this represents the potential control-flow or data-flow path through which the vulnerability could be activated.
  It records how input values propagate to sensitive operations 
  (e.g., ``user input $\rightarrow$ array index $\rightarrow$ buffer write''). 
  Trigger paths are generated by analyzing the candidate reports and constructing symbolic traces across the code under analysis.
\end{itemize}

By structuring hypotheses in this way, VulAgent creates a foundation for systematic validation in the following phase, where hypotheses  can be checked against.

In order to construct vulnerability hypotheses, we design a prompt that guides the model to organize candidate issues into structured hypotheses.
The TriggerPlannerAgent prompt is designed to guide the agent in constructing 
feasible execution paths for each candidate vulnerability. Unlike other prompts 
that focus on detection patterns, this prompt instructs the agent to: 

\begin{enumerate}
  \item identify the sink (such as memory write, dereference, or system call) reported by upstream analyzers, 
  \item backtrack through the function code to determine how data flows into this sink, 
  \item trace attacker-controllable inputs (parameters, buffers, file reads, deserialized structures) along control-flow and data-flow edges until they reach the sink, and 
  \item record intermediate steps (assignments, transformations, conditionals) as part of the execution trace. 
\end{enumerate}

\begin{wrapfigure}{l}{0.48\textwidth} 
    \centering
    \vspace{-10pt} 
    \includegraphics[width=0.48\textwidth]{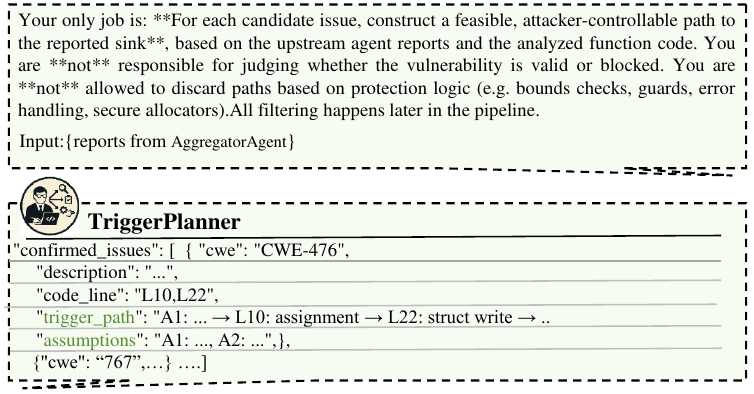}
    \label{fig:triggerplanner}
    \vspace{-10pt} 
\end{wrapfigure}
The prompt emphasizes that the agent must always attempt path construction if the 
sink is syntactically present, and must not filter or discard paths based on guards, 
bounds checks, or error handling. 
Instead, any hypotheses conditionals are to be explicitly annotated as part of the trace. 
This design ensures that the output is a structured vulnerability hypothesis, consisting of a trigger path annotated with conditionals, while leaving all pruning and validity judgments to downstream agents.

For each aggregated entry $\tilde t_{i}^{(k)} \in \widetilde{T}_i$ with predicted CWE type $\mathrm{cwe}_{i,k}$ and reported sink $s_{i,k}$, the TriggerPlannerAgent constructs a structured hypothesis

The TriggerPlannerAgent deterministically maps each candidate to a structured hypothesis:
\begin{equation}
  h_{i,k} \;=\; \Psi\!\bigl(\tilde t_{i}^{(k)},C_i \bigr)
  \;=\; \bigl(\,\mathrm{cwe}_{i,k},\; \mathcal{A}_{i,k},\; \mathcal{P}_{i,k}\,\bigr),
  \quad k=1,\dots,M_i ,
\end{equation}
where $\mathcal{A}_{i,k}$ is the set of \emph{assumption conditions} (e.g., ``$|x|>\!n$'', ``pointer non-NULL''),
and $\mathcal{P}_{i,k}$ is the \emph{trigger path} from an attacker-controllable source to the sink $s_{i,k}$.

We model a trigger path as a labeled path over program graphs:
\begin{equation}
  \mathcal{P}_{i,k} \;=\; \bigl\langle
    v_1 \xrightarrow{e_1} v_2 \xrightarrow{e_2} \cdots \xrightarrow{e_{m-1}} v_m
  \bigr\rangle ,
  \quad 
  v_1 \in \mathcal{S}^{\text{att}}_{i,k},\;\; v_m = s_{i,k},
\end{equation}
where $\mathcal{S}^{\text{att}}_{i,k}$ is the set of attacker-controllable sources
(e.g., parameters, file/network reads, deserialized inputs), and each edge $e_j$ is
either a control- or data-dependence edge. All guard conditions encountered along
the path are recorded into $\mathcal{A}_{i,k}$ instead of being used to prune the path.




In conclusion, the hypothesis construction step augments each vulnerability entry in the structured CWE report with two attributes—assumption conditions and a trigger path—which then serve as the basis for subsequent validation.

\subsection{Phase III: Hypothesis-Conditions Validation}


In this phase, VulAgent verifies the \textit{hypothesis conditions} extracted from each aggregated report.
Unlike prior methods that attach raw context, we turn each condition into a targeted query, checking its
consistency against the program context. This focuses on the feasibility of conditions themselves rather
than prematurely reasoning about the execution path.

Let $C_i^{\mathrm{ctx}}$ denote the optional surrounding
context (e.g., callee bodies, globals, types, invariants). For a hypothesis
$h_{i,k} = \bigl(\mathrm{cwe}_{i,k}, \mathcal{A}_{i,k}, \mathcal{P}_{i,k}\bigr)$
derived from $\tilde t_{i}^{(k)}$, we validate each condition
$a_{i,k,j} \in \mathcal{A}_{i,k}$ via a per-condition validation function:
\begin{equation}
  (\nu_{i,k,j}, \,\mathcal{E}_{i,k,j}) \;=\; f_{\mathrm{va}}\!\bigl(a_{i,k,j},\, c_i,\, C_i^{\mathrm{ctx}}\bigr),
\end{equation}
where $\nu_{i,k,j} \in \{\mathrm{valid},\, \mathrm{contradicted},\, \mathrm{plausible}\}$ is the decision
for condition $a_{i,k,j}$, and $\mathcal{E}_{i,k,j}$ collects supporting evidence
(e.g., cited lines, inferred invariants, constant ranges).

We summarize the validated subset as
\begin{equation}
  \widehat{\mathcal{A}}_{i,k} \;=\; \bigl\{\, a_{i,k,j} \in \mathcal{A}_{i,k} \;\big|\; \nu_{i,k,j}=\mathrm{valid,plausible} \,\bigr\}.
\end{equation}
Only $\widehat{\mathcal{A}}_{i,k}$ proceeds to later stages; path reasoning is deferred until the trigger-path validation phase.

The AssumptionPruner Agent prompt is designed to guide the agent to confirm whether the assumed conditions for triggering each build are reasonable.
%
The core principle of its prompt is:

  (1) extract every assumption explicitly recorded in the trigger path, 
  (2) classify each assumption as plausible, contradicted, or unknown, 
  (3) identify whether invalidating a contradicted assumption would cause the entire trigger path to break, and 
  (4) record structured reasoning for each classification decision. 

\begin{wrapfigure}{l}{0.48\textwidth} 
    \centering
    \vspace{-10pt} 
    \includegraphics[width=0.48\textwidth]{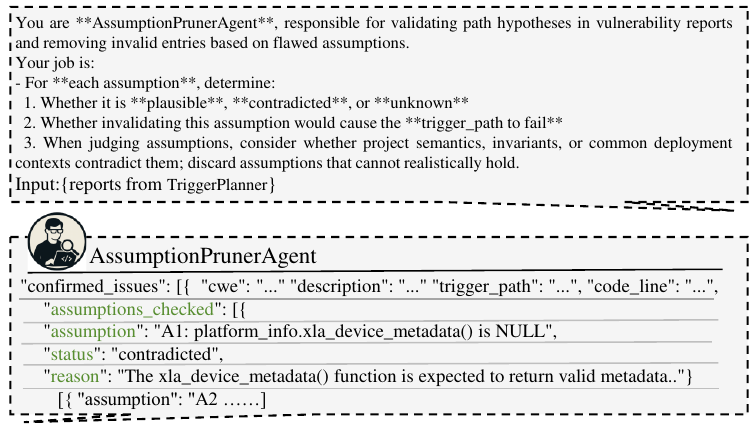}
    \label{fig:memorylayoutagent}
    \vspace{-20pt} 
\end{wrapfigure}
A  challenge in hypothesis validation is that many vulnerability hypotheses rely on interactions between control-flow, data-flow, and inter-procedural calls.
For instance, issues such as illegal pointer dereferences, unchecked user inputs, or misuse of sensitive resources cannot be validated by inspecting a single code block in isolation.
They typically emerge from how values propagate along data dependencies, how conditions constrain execution paths, and how functions invoke one another.
Without these contexts $C_i^{\mathrm{ctx}}$, LLM-based reasoning tends to remain speculative, leading to false alarms.

\begin{wrapfigure}{l}{0.38\textwidth} 
    \centering
    \includegraphics[width=0.38\textwidth]{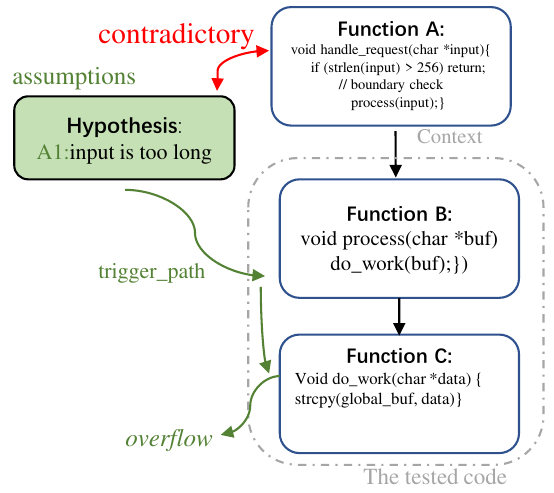}
    \caption{Code Context–driven hypothesis validation.}
     \label{fig:hypothesis_validation}
\end{wrapfigure}
To solve this problem, our hypothesis validation agent utilized program structure information extracted by static tools such as Joern~\cite{JoernTool}.
For the PrimeVul dataset, we constructed contextual information using Joern. 
For SVEN, we developed a static tool that traces back to the original file through the submission link to build minimal contextual information, including the project name, functions defined or called within the file, referenced libraries.

Specifically, Joern was consulted to obtain the following information:

(i) control-flow graphs (CFGs) that capture feasible execution paths;

(ii) data-flow dependencies showing how inputs and variables propagate;

(iii) call graphs that uncover inter-function interactions.


Figure~\ref{fig:hypothesis_validation} illustrates hypothesis validation from a context-first perspective.
The call graph reveals the surrounding callers (Function A and Function B) of the tested function Function C that contains the sink (\texttt{strcpy}).
These callers are collected as context.
Given the hypothesis A1 (“input is too long”), the AssumptionPruner checks whether this assumption is consistent with the collected context.
In Function A, an input-length check (e.g., \texttt{if (strlen(input) > 256) return;}) verifies and rejects overlong input before the value is forwarded to Function C.
This evidence conflicts with A1, so the assumption is classified as \emph{contradicted}; because the hypothesis relies on A1, the hypothesis is rejected.

Overall, rather than applying generic heuristics or passively concatenating surrounding code, we use the hypothesis conditions to steer how the agent retrieves and consumes program context. 
For each assumption, the AssumptionPruner consults code context to seek confirming or refuting evidence; contradictions to structural invariants, API semantics, lifecycle guarantees, or synchronization rules are marked \textit{contradicted}, while hypotheses  lacking sufficient evidence remain \textit{plausible} but flagged.

\subsection{Phase IV: Hypothesis-Path Verification}

In this phase, VulAgent examines whether the validated hypotheses can still be triggered in practice, by checking for existing defensive mechanisms along the execution paths. 
To be specific, the assumption conditions have already been 
validated, so the task is to analyze the \textit{trigger path} and detect guards that may neutralize the vulnerability (e.g., boundary checks, 
null checks, early returns, or structured error handling). This prompt acts as the final checkpoint that decides whether each hypothesis’s trigger path remains valid under the given program context. So the core requirements of the prompt words are as follows:

\begin{wrapfigure}{l}{0.48\textwidth} 
    \centering
    \vspace{-10pt} 
    \includegraphics[width=0.48\textwidth]{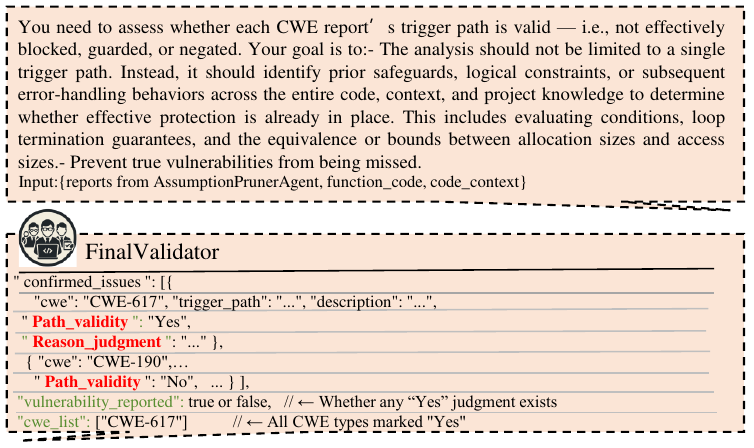}
    \label{fig:final_validator}
    \vspace{-10pt} 
\end{wrapfigure}

\begin{enumerate}
  \item make a binary decision on the trigger path under context, using the guard-dominance idea: treat the path as valid unless a protection placed before the sink blocks all feasible routes;
  \item consult control flow, data flow, call relationships, and semantic context (API behavior, lifecycle, configuration) to seek concrete evidence;
  \item reject the path when clear pre-sink defenses, sanitization/overwrite, unmet premises, protective API semantics, unreachability, or harmless-only outcomes are evidenced; otherwise retain the path, and forward uncertain cases without suppression but with a brief rationale;
  \item return a strict JSON verdict per CWE entry, along with an overall vulnerability flag and the list of CWEs judged valid.
\end{enumerate}



Given a validated hypothesis
$h_{i,k} = \bigl(\mathrm{cwe}_{i,k},\\ \widehat{\mathcal{A}}_{i,k}, \mathcal{P}_{i,k}\bigr)$,
the line-numbered function under test $c_i$, and an optional program context
$C_i^{\mathrm{ctx}}$ (default uses),
we decide whether effective defenses dominate all feasible routes to the sink
under the validated assumptions. The defensive verification function returns a
binary decision on the trigger path:
\begin{equation}
 \hspace*{1.02\linewidth}
  r_{i,k} \;=\; f_{\mathrm{pc}}\!\bigl(\mathcal{P}_{i,k},\, c_i,\, C_i^{\mathrm{ctx}};\, \widehat{\mathcal{A}}_{i,k}\bigr).
\end{equation}
Here $r_{i,k}$ is a binary decision (\emph{retained} or \emph{discarded}).
A hypothesis is marked \emph{discarded} iff its trigger path $\mathcal{P}_{i,k}$
is \emph{dominated} by protections that (i) precede the sink and (ii) cover
\emph{all} feasible routes from attacker-controllable sources to the sink
given $\widehat{\mathcal{A}}_{i,k}$ (e.g., length/null/capability checks gating
the sink). Otherwise, the hypothesis is \emph{retained}.


\section{STUDY SETUP}
\subsection{Datasets}
To rigorously assess effectiveness and ensure comparability,  we adopt two widely-used, high-quality datasets: \textbf{PrimeVul}~\cite{ding2025vulnerability} and \textbf{SVEN}~\cite{he2023large}.

\paragraph{PrimeVul.}
PrimeVul is a C/C++ vulnerability dataset collected from real-world open-source projects via a data-labeling procedure, covering $6{,}968$ vulnerable samples across $6{,}827$ commits and $755$ projects, spanning $140$ CWE categories~\cite{ding2025vulnerability}.
In our experiments, we follow the same evaluation split and baseline protocols as the compared work to ensure fair comparison~\cite{widyasari2025let}. Under this setting, the validation set consists of 435 pairs of code, with each pair containing the version before the fix and the version after the fix.

\paragraph{SVEN}
SVEN is a hand-checked, function-level corpus built from security-fixing commits, formed by filtering and pairing samples from CrossVul, Big-Vul, and VUDENC to emphasize data quality. It covers nine CWEs and includes 1,606 programs (803 vulnerable–fixed pairs) across C/C++ and Python, with a 9:1 train/validation split~\cite{he2023large}.



\subsection{Experimental Details}
To ensure reproducibility, all large language models used in this work were accessed via commercial APIs, 
with temperature set to 0 and top\_p set to 1. All datasets follow publicly available splits. 
Prompt construction does not include any few-shot code examples.
%
We have also released all generated vulnerability reports, the prompts and system logic used in VulAgent, 
the static tool for constructing context for the SVEN dataset, and the resulting context-augmented dataset package.
To control cost and mitigate the latency of reasoning-enabled models, we default to non-reasoning models and enforce a strict JSON-only output format to reduce token usage. We empirically sampled our runs and found that, on average, each sample issues 8.83 calls and generates 7{,}036.95 output tokens. The system adopts a multi-threaded design with a concurrency of 8 to enhance the detection efficiency.



\subsection{Performance Metrics}
\textbf{Usability Evaluation.}
To faithfully evaluate the effectiveness of vulnerability detection in the vulnerability–fix pair setting, we adopt an evaluation protocol consistent with prior work~\cite{wen2025boosting}, while additionally incorporating False Positive Rate (FPR) to better assess the practical usability of the models.

Specifically, we report the following:  
\textbf{Accuracy (ACC)}. Accuracy reflects the proportion of correctly predicted samples among all functions, offering a global view of the model’s predictive capability:
\begin{equation}
Accuracy = \frac{TP + TN}{TP + TN + FN + FP}.
\end{equation}
When combined with the False Positive Rate (FPR), ACC provides a more practical perspective on usability: a model with high accuracy but also a high false positive rate may still be unsuitable for real-world deployment.  

\textbf{F1-score}. F1 is the harmonic mean of precision and recall:
\begin{equation}
F1 = \frac{2 \times \text{Precision} \times \text{Recall}}{\text{Precision} + \text{Recall}}.
\end{equation}
While F1 is widely used to balance false negatives and false positives, in vulnerability detection it may present inflated values when the false positive rate is high~\cite{steenhoek2022deepdfa}.  
Therefore, we additionally report FPR to complement F1 and better reflect the practical applicability of detection models.

\textbf{False Positive Rate (FPR)}: FPR is the fraction of benign functions that are incorrectly flagged as vulnerable:
\begin{equation}
FPR = \frac{FP}{FP + TN}
\end{equation}
A low FPR is crucial in practice, as it reduces unnecessary manual reviews and improves the usability of vulnerability detection systems in real-world settings.

\textbf{Understanding Evaluation.}
To further assess whether a model can truly distinguish between vulnerable and fixed versions of code, we use pair-wise evaluation metrics~\cite{ding2025vulnerability}:

Pair-wise Correct Prediction (P-C): the percentage of function pairs in which both the vulnerable function and its fixed counterpart are correctly classified. It reflects the model’s ability to capture fine-grained semantic differences.

Pair-wise Reversed Prediction (P-R): the percentage of pairs where both functions are incorrectly classified as vulnerable.

Pair-wise Score (VPS)~\cite{wen2025boosting}:
Since P-C and P-R are defined as percentages of total pairs, VPS is computed as their difference, reflecting the net gain of correct over reversed predictions.
A higher VPS means the model better captures subtle differences between vulnerable and fixed code, indicating stronger vulnerability understanding ability.

In summary, these metrics provide a balanced evaluation framework that aligns with our method’s goal of improving real-world applicability through reduced false positives.

\subsection{Baseline Methods}
For the single-agent baseline, we adopt the chain-of-thought (CoT) prompting strategy~\cite{ding2025vulnerability}, which has been widely explored in prior work for code reasoning tasks and achieves strong results on PrimeVul~\cite{ding2025vulnerability}. CoT prompting encourages the model to reason step-by-step when evaluating potential vulnerabilities.

For the multi-agent baselines, we include three recent representative approaches. The first is GPTLens\cite{hu2023large}, which employs two agents: an auditor agent that proposes potential vulnerabilities and a critic agent that scores them, with the final decision based on the critic’s score.

The second is VulTrial\cite{widyasari2025let}, which follows a courtroom-inspired setting with four role-specific agents (security researcher, code author, judge, and review board). Unlike GPTLens, where the critic agent determines the final outcome, VulTrial leverages a neutral review board agent for the final decision.

The third is LLM-MultiRole~\cite{mao2024multi}, which introduces a multi-role setting where LLM agents take on roles such as developers and testers, simulating a real-world code review process through collaborative discussion.

\textbf{Models.}
To evaluate the generality of our approach across different large language models, we conduct experiments on three representative and up-to-date models.

\textbf{Qwen3-235B-A22-instruct~\cite{qwen3technicalreport}.} 
Qwen3 is a state-of-the-art open-source model family, and we employ its 235B parameter instruction-tuned variant, which has demonstrated strong reasoning and code understanding capabilities. Its large scale and instruction alignment make it well-suited for security-related code analysis tasks.

\textbf{GPT-4o~\cite{openai2023gpt4}.} 
GPT-4o  is OpenAI’s flagship model, designed as a successor to GPT-4. It provides comparable or better accuracy while significantly improving efficiency, and supports a wide context window (128k input tokens and 16k output tokens), enabling it to capture long code contexts.

\textbf{DeepSeek v3.1~\cite{deepseekai2024deepseekv3technicalreport}.} 
DeepSeek v3.1 is a recent large-scale model with competitive reasoning and multilingual coding capabilities. It has been widely recognized for strong performance in code-related tasks and provides a representative benchmark beyond the OpenAI and Qwen ecosystems.

In addition, we also include \textbf{GPT-3.5} in our baseline comparison. For this model, we directly cite results from VulTrial~\cite{widyasari2025let}.
We thank the authors for releasing their results, but since GPT-3.5 is now considered outdated, we do not further evaluate our own method on this model. 
Furthermore, we also report results of our fine-tuned method on UniXcoder~\cite{guo2022unixcoder}, which has been recognized as one of the strongest code pre-trained models for vulnerability detection on the CodeXGLUE benchmark~\cite{lu2021codexglue}. 
All methods have made every effort to utilize their open-source implementations and results.
Among all the methods with context requirements, this article ensures that completely consistent context information is provided.


\section{EXPERIMENTS}
In this section, we conduct extensive experiments to demonstrate 
the effectiveness of our proposed approach and to analyze the 
factors contributing to its performance. Specifically, we aim to 
answer the following research questions:

\textbf{RQ1:} How effective is \textit{VulAgent} compared with state-of-the-art?

\textbf{RQ2:} How well does \textit{VulAgent} generalize across different LLMs?

\textbf{RQ3:} What impact would the lack of relevant context have on the performance of \textit{VulAgent}?

\textbf{RQ4:} How does VulAgent perform across different types of vulnerabilities (CWE categories)?

\textbf{RQ5:} How effective is our proposed hypothesis validation process?

\subsection{RQ1. Effectiveness of VulAgent }

\textbf{Setup.} 
We evaluated VulAgent against state-of-the-art LLM-based baselines on two representative datasets, PrimeVul and SVEN. 
For fairness, all baseline results are reproduced following their original settings, while VulAgent is evaluated using the proposed multi-agent pipeline with hypothesis validation. 

\textbf{Overall Results.} 
Table~\ref{tab:pairwise_results} summarizes the comparison. 

\begin{table*}
\centering
\caption{Comparison of vulnerability detection performance across two datasets. }
\vspace{-1em}
\label{tab:pairwise_results}
\resizebox{\textwidth}{!}{
\begin{tabular}{l l l c c c c c c|  c c c  c c c}
\toprule
\multirow{2}{*}{Model} & \multirow{2}{*}{Agents} & \multirow{2}{*}{Method} 
& \multicolumn{6}{c|}{PrimeVul} 
& \multicolumn{6}{c}{SVEN} \\
\cmidrule(lr){4-9} \cmidrule(lr){10-15}
 &  &  & P-C$\uparrow$ & P-R$\downarrow$ & VPS$\uparrow$ & FPR$\downarrow$& F1$\uparrow$ & ACC$\uparrow$ 
 & P-C$\uparrow$ & P-R$\downarrow$ & VPS$\uparrow$ & FPR$\downarrow$& F1$\uparrow$ & ACC$\uparrow$ \\

 \midrule
\multirow{1}{*}{UniXcoder} 
& - & SFT
& 7.62 & 0.92& 6.69& 22.9& 54.0&  53.4
& 13.2& 0.00 & 13.2& 1.20& 25.0&  56.6\\

\midrule
\multirow{3}{*}{GPT-3.5} 
& Single-Agent & CoT 
& 6.21 & 5.50 & 0.71 & 10.3 &18.1 &  50.3
& 1.20 & 0.00 & 1.20 & 95.3 &66.1 & 51.8\\
& Multi-Agent & GPTLens
& 10.1 & 6.44 & -0.91& 94.7 & 65.0& 49.6
& 6.00 &7.60 & -1.2 & 19.3& 26.3& 49.4 \\
& Multi-Agent & VulTrial 
& 18.6 & 11.4& 1.38 & 23.4 & 33.4& 50.7
& 1.20& 1.20 & 0.00& 97.6 & 65.8& 50.0 \\
\midrule
\multirow{3}{*}{GPT-4o} 
& Single-Agent & CoT 
& 9.20 & 6.67 & 2.50 & 30.8 & 28.1& 51.7
& 1.20 & 2.40 & -1.20 & 97.5 &65.5&  49.4 \\

& Multi-Agent & LLM-MultiRole
& 1.61 & 6.44 & -5.05& 17.3 & 65.7& 47.3
& 3.61 & 12.0 & -8.43 & 15.6 & 11.7& 45.7 \\

& Multi-Agent & GPTLens
& 10.1 & 6.76& 3.60 & 61.9 &18.8&  50.2
& 3.61 & 3.61 & 0.00 & 92.8 & 64.9& 50.2 \\
& Multi-Agent & VulTrial 
& 18.6 & 11.4 & 7.13 & 52.6 &56.1&  53.4
& 4.82 & 1.20 & 3.60 & 95.2 &67.2&  52.0 \\
\hline
& Multi-Agent & \textbf{VulAgent} 
& \textbf{26.6} & 10.1 & \textbf{16.5} & \textbf{36.7} &56.2 & \textbf{58.4}
 & \textbf{26.5} & 6.02& \textbf{20.4} & 55.4 & 65.62& \textbf{60.2} \\
\bottomrule
\end{tabular}}
\vspace{-1em}
\end{table*}

\begin{figure}[t]
  \centering
  \includegraphics[width=0.45\linewidth]{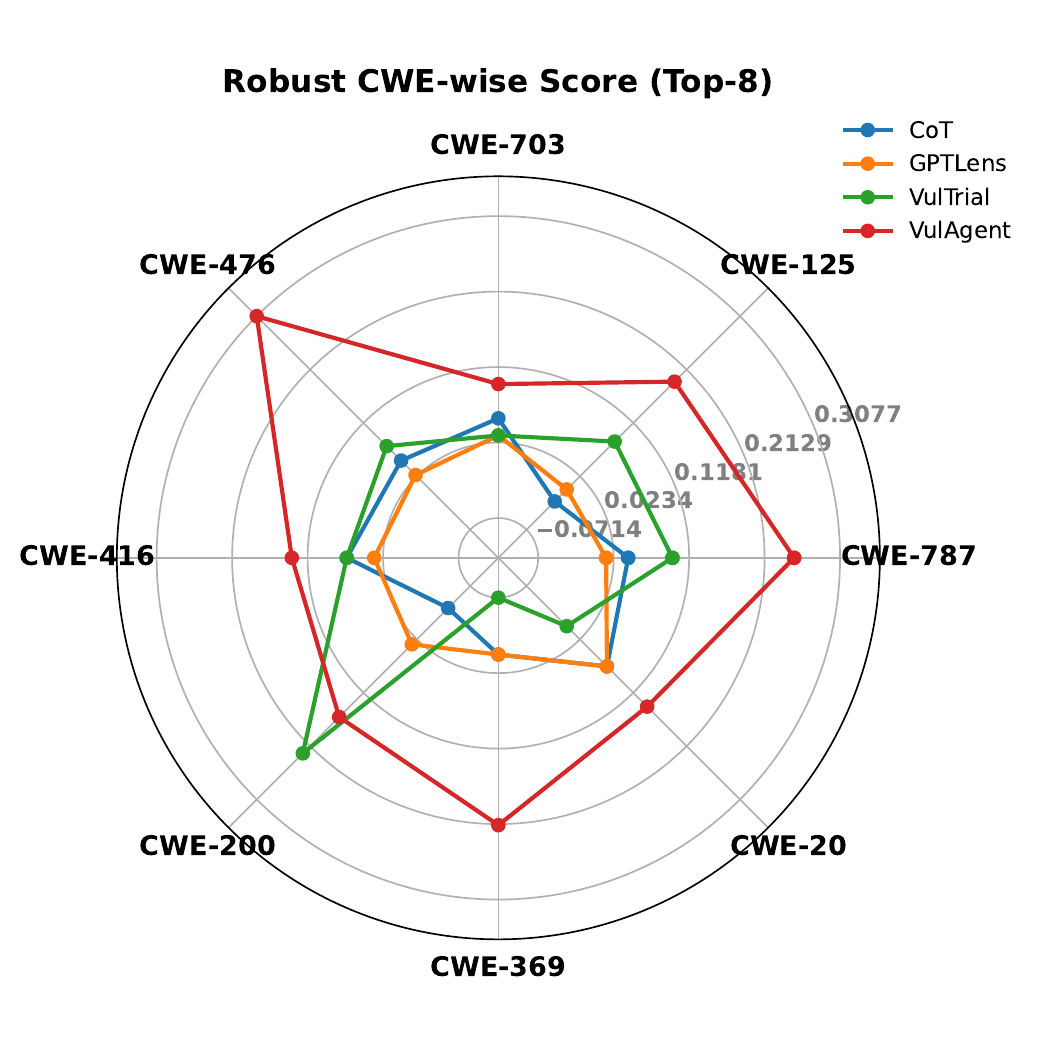}
  \includegraphics[width=0.45\linewidth]{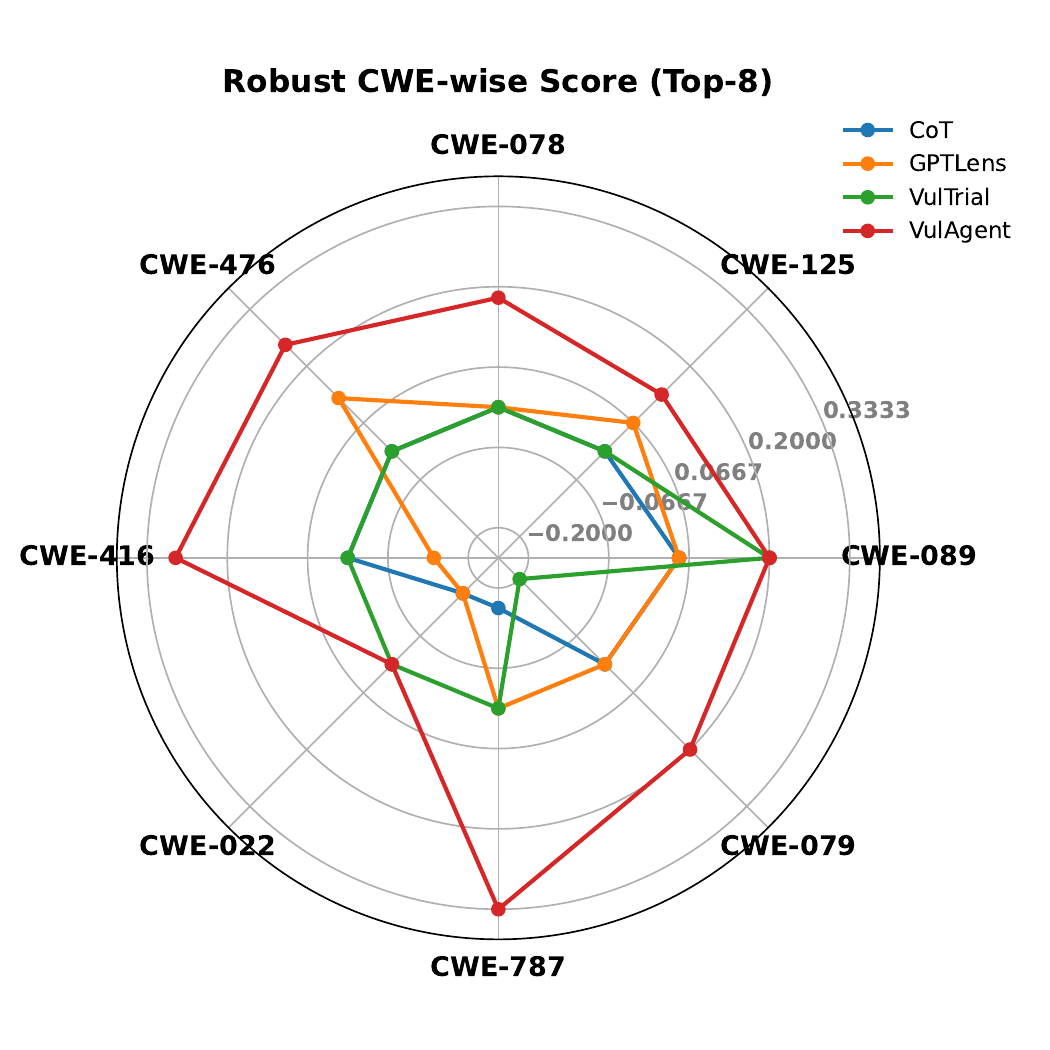}
 \caption{Performance of different methods across the Top-8 CWE categories on PrimeVul (left) and SVEN (right). 
The score is defined as $(\#\text{correct}-\#\text{wrong})/\#\text{total}$ for each CWE.}
  \label{fig:cwe_radar_pri}
  \label{fig:cwe_radar_seven}
  \vspace{-1em}
\end{figure}

On PrimeVul, VulAgent achieves a P-C of 26.6\%, significantly surpassing VulTrial (18.6\%, +43.0\%). 
At the same time, FPR is reduced from 52.6\% to 36.8\% (a relative reduction of 30.0\%), and accuracy improves from 53.4\% to 58.4\% (+5.0 percentage points). 
These results highlight that the hypothesis-validation mechanism effectively exploits contextual information to prune false positives while improving accuracy.  

On SVEN, where most baselines show severe degradation, VulAgent still maintains clear advantages. 
It achieves a P-C of 26.5\%, compared to VulTrial’s 4.82\% (+449.8\%). 
FPR decreases from 95.2\% to 55.4\% (a relative reduction of 41.8\%), while accuracy rises from 52.0\% to 60.2\% (+8.2 percentage points). 
Driven by high recall yet accompanied by substantial false-positive rates (FPR), several baselines maintain respectable F1 scores despite low overall accuracy.
This demonstrates that VulAgent remains robust even under limited-context settings, validating its generality across datasets.  

To further illustrate the robustness of different methods across vulnerability categories, 
we plot the robust CWE-wise scores on the PrimeVul and SVEN datasets 
( as shown in Figure~\ref{fig:cwe_radar_pri}). 
The robust score is defined as $(\#\text{correct}-\#\text{wrong})/\#\text{total}$ for each CWE, 
ranging from $-1$ to $1$. 
Negative scores indicate that mispredictions outnumber correct predictions, 
highlighting instability of the method in that category. 
As shown in the radar charts, \textit{VulAgent} consistently achieves 
higher and more balanced scores across the Top-8 CWE types in both datasets. 
In PrimeVul (Figure~\ref{fig:cwe_radar_pri}  (left)), \textit{VulAgent} outperforms baselines 
particularly on CWE-703, CWE-476, and CWE-416. 
We note that the performance gap on CWE-200 is less pronounced, and we provide a detailed analysis of its limitations in RQ3.

In SVEN (Figure~\ref{fig:cwe_radar_seven}), the advantage is even clearer: 
while VulTrial and other baselines fail on categories such as CWE-089 and CWE-078, 
\textit{VulAgent} maintains stable positive scores across all eight categories. 
These results confirm that our hypothesis validation process not only 
improves overall metrics but also enhances robustness across diverse vulnerability classes.

We further examined the failure cases of baselines. 
Single-agent and reflection-based methods (e.g., GPTLens, VulTrial) tend to over-approximate risks, leading to excessive false positives (FPR $>$ 90\% on SVEN). 
VulAgent, by contrast, filters out such speculative paths through its two-stage validation, which explains the large improvement in P-C and the substantial reduction in FPR. 
These findings indicate that our design not only increases detection accuracy but also lowers verification costs by reducing the burden of manual false-positive review.

\subsection{RQ2: Generalization Ability of VulAgent}

\begin{table*}
\centering
\caption{Comparison of model generalization performance across two datasets.}
\label{tab:rq2_generalization}
\resizebox{\textwidth}{!}{
\begin{tabular}{l c c c c cc  | c c cc c c}
\toprule
\multirow{2}{*}{Model} 
& \multicolumn{6}{c|}{PrimeVul} 
& \multicolumn{6}{c}{SVEN} \\
\cmidrule(lr){2-7} \cmidrule(lr){8-13}
 & P-C$\uparrow$ & P-R$\downarrow$ & VPS$\uparrow$ & FPR$\downarrow$ & F1$\uparrow$& ACC$\uparrow$ 
 & P-C$\uparrow$ & P-R$\downarrow$ & VPS$\uparrow$ & FPR$\downarrow$ & F1$\uparrow$& ACC$\uparrow$ \\
\midrule
DeepSeek v3.1  
& 17.70 & \textbf{8.74} & 8.96 & \textbf{19.95} &41.59 &54.73

& 30.12 & 9.64 & 20.48 & 31.30 & 56.58&60.20 \\
Qwen3-235B-A22-instruct     
& 25.75 & 11.03 & 14.71 & 40.15 & \textbf{57.70}&57.36

& \textbf{30.12} & 10.84 & 19.28 & \textbf{28.91} &54.42 &59.63\\
GPT-4o         
& \textbf{26.67} & 10.11 & \textbf{16.55} & 36.78 &56.11 &\textbf{58.62}
 & 26.51 & \textbf{6.02} & \textbf{20.49} & 55.42 & \textbf{65.62}&\textbf{60.20} \\

\bottomrule
\end{tabular}}
\end{table*}

To evaluate the generalization capacity of our proposed VulAgent framework, we assess its vulnerability prediction performance across two datasets: PrimeVul and SVEN. Table~\ref{tab:rq2_generalization} presents the results.

Across both datasets, GPT-4o achieves the highest overall accuracy, with 58.62\% on PrimeVul and 60.20\% on SVEN, demonstrating the strongest generalization ability. Notably, it maintains a high pair score while balancing false positives reasonably well.

In contrast, DeepSeek v3.1 consistently reports the lowest false positive rates—19.95\% on PrimeVul and 31.30\% on SVEN—indicating a conservative detection style. Although its P-C and VPS are lower than GPT-4o and Qwen3-235B, this restraint makes it suitable for high-precision scenarios where over-reporting is costly.

\textbf{Summary.}  
These results confirm that VulAgent generalizes robustly across different LLM backbones: 
it consistently improves accuracy and reduces false positives regardless of whether the underlying model 
prioritizes precision (DeepSeek), balance (Qwen3), or overall accuracy (GPT-4o).

\subsection{RQ3. Effects of Code Context for VulAgent }

\begin{table}
\centering
\caption{Effect of code context on VulAgent (ChatGPT-4o backbone).}
\label{tab:rq3_results}
\footnotesize
\setlength{\tabcolsep}{15pt}
\begin{tabular}{lcccccc}
\toprule
Dataset / Setting & P-C & P-R & VPS & FPR & F1 & ACC \\
\midrule
PrimeVul (w/o context) & 17.50 & 7.50 & 10.00 & 38.75&52.00 & 55.00 \\
PrimeVul (w/ context)  & \textbf{28.73} &\textbf{ 10.70} & \textbf{18.03} & \textbf{36.34} &\textbf{57.02} &\textbf{59.01} \\ \hline
SVEN (w/o context)     & 20.48 & 7.20 & 13.25 & 60.00 & 63.27&56.63 \\
SVEN (w/ context)      & \textbf{26.51} & \textbf{6.02} & \textbf{20.48} &\textbf{55.42} &\textbf{ 65.62}& \textbf{60.20} \\
\bottomrule
\end{tabular}
\vspace{-1em}
\end{table}

To investigate the role of code context, we conduct experiments with and without context on both datasets, and the results are shown in Table~\ref{tab:rq3_results}. 

For \textbf{PrimeVul}, we followed prior work and leveraged its annotated contextual information. Among the 435 code pairs in the test set, 304 pairs have available context.
We report results separately for the context-available subset and the context-missing subset. The results show that when context is incorporated, VulAgent achieves a pair-wise correct prediction (P-C) of 28.73\%, compared to 17.50\% without context, yielding a relative improvement of 64.1\%. VPS also increases from 10.00 to 18.03, while accuracy rises from 55.00\% to 59.01\%. These results confirm that our hypothesis-validation mechanism can effectively exploit contextual information to enhance reasoning and reduce false positives.  

For \textbf{SVEN}, the original dataset does not provide contextual annotations. To enable comparison, we employed static analysis tools to retrieve project-level information, including related function calls, imports, and project metadata, and integrated these into VulAgent. The results demonstrate similar improvements: P-C increases from 20.48\% to 26.51\% (+29.4\% relative), VPS improves from 13.25 to 20.48, and accuracy rises from 56.63\% to 60.20\%.  

\textbf{Summary.}  
Overall, these results highlight that contextual information substantially improves VulAgent’s effectiveness across different datasets. By coupling context retrieval with hypothesis validation, VulAgent avoids the naïve concatenation strategy and instead leverages context in a targeted and principled manner.  


\subsection{RQ4. Effectiveness Across Different CWE Categories}
\begin{table*}[ht]
\centering
\caption{Performance of VulAgent across different CWE categories on PrimeVul (Top-8) and SVEN (All).}
\vspace{-1em}
\label{tab:cwe_results}
\footnotesize
\setlength{\tabcolsep}{18pt}
\begin{tabular}{l lccccc}
\toprule
Dataset & CWE & Prec & Rec & F1 & Acc & FPR \\
\midrule
\multirow{5}{*}{PrimeVul} 

& CWE-476 & 0.688 & 0.564 & 0.620 & 0.654 & 0.256 \\
& CWE-787 & 0.661 & 0.514 & 0.578 & 0.625 & 0.264 \\
& CWE-369 & 0.714 & 0.357 & 0.476 & 0.607 & 0.143 \\
& CWE-125 & 0.610 & 0.532 & 0.568 & 0.596 & 0.340 \\
& CWE-200 & 0.588 & 0.625 & 0.606 & 0.581 & 0.467 \\
& CWE-416 & 0.583 & 0.483 & 0.528 & 0.569 & 0.345 \\
& CWE-20  & 0.571 & 0.571 & 0.571 & 0.571 & 0.429 \\
& CWE-703 & 0.558 & 0.511 & 0.533 & 0.548 & 0.413 \\
\midrule
\multirow{9}{*}{SVEN} 
& CWE-190 & 0.667 & 0.800 & 0.727 & 0.700 & 0.400 \\
& CWE-787 & 0.667 & 0.667 & 0.667 & 0.667 & 0.333 \\
& CWE-416 & 0.667 & 0.571 & 0.615 & 0.643 & 0.286 \\
& CWE-476 & 0.583 & 0.875 & 0.700 & 0.625 & 0.625 \\
& CWE-089 & 0.556 & 1.000 & 0.714 & 0.600 & 0.800 \\
& CWE-079 & 0.667 & 0.400 & 0.500 & 0.600 & 0.200 \\
& CWE-078 & 0.583 & 0.636 & 0.609 & 0.591 & 0.455 \\
& CWE-125 & 0.550 & 0.733 & 0.629 & 0.567 & 0.600 \\
& CWE-022 & 0.500 & 0.667 & 0.571 & 0.500 & 0.667 \\
\bottomrule
\end{tabular}
\vspace{-1em}
\end{table*}
Table~\ref{tab:cwe_results} reports VulAgent’s performance across representative CWE categories on PrimeVul (Top-8) and SVEN (all). The results reveal distinct patterns depending on whether vulnerabilities are easier to hypothesize and how strongly they depend on contextual information.

\textbf{(1) Easy-to-hypothesize + weak context dependency.}  
Errors such as CWE-476 (null dereference) and CWE-190 (integer overflow) 
are easy to hypothesize because any pointer dereference may be imagined as operating on a null value, 
and any arithmetic on attacker-influenced operands may overflow. 
Their validation usually requires only local checks, such as null guards or range bounds, 
so once hypothesized they can often be confirmed or refuted directly within the function.
This makes such vulnerabilities easier to confirm compared with other categories.
At the same time, because these patterns are so easily hypothesized, 
they also tend to generate false positives when the actual protection comes 
from global invariants or project-level contracts rather than local checks. 
This explains why VulAgent achieves high effectiveness in confirming real cases, 
while still suffering from relatively high false positive rates in some projects.

\textbf{(2) Easy-to-hypothesize + strong context dependency.}  
Vulnerabilities such as CWE-089 (SQL injection), CWE-125 (out-of-bounds read), 
and CWE-022 (path traversal) are easy to hypothesize because their syntactic patterns 
immediately suggest attacker-controlled risk: any dynamic SQL concatenation may be imagined as injectable, any array access may be imagined as out-of-bounds, and any file path operation may be imagined as  subject to traversal. 
However, confirming whether these hypotheses are actually exploitable depends on 
global project constraints---for instance, whether SQL statements are parameterized, whether array indices are bounded by upstream logic, or whether the file system layer enforces path checks. 
Such protections are often outside the local context scope, so they cannot be reliably observed without richer contextual information. 
As a result, VulAgent shows a consistent pattern in these categories: 
recall remains high because the model tends to flag nearly all candidate cases, 
but false positive rates are also high because many of these hypotheses cannot be overturned 
given only limited context. 
This reflects that while such flaws are easy to hypothesize, 
they are intrinsically hard to validate without full-system semantic knowledge.

\textbf{(3) Hard-to-hypothesize + weak context dependency.}  
Semantic issues such as CWE-703 (improper handling of exceptional conditions) 
are difficult to hypothesize because their triggering conditions are not syntactically obvious. 
Unlike memory or injection flaws, these vulnerabilities only manifest if execution enters 
rare or unusual error-handling branches, which requires the model to speculate about 
uncommon states that are rarely visible in local code. 
Once hypothesized, validation 
would in principle require only local checks (e.g., whether an error is properly propagated), but the main challenge is that such hypotheses are seldom constructed in the first place. 
As a result, VulAgent reports relatively few false positives in this category, 
but its recall and overall accuracy remain low, reflecting that these flaws are both 
hard to hypothesize and hard to detect reliably.

\textbf{(4) Hard-to-hypothesize + strong context dependency.}  
Vulnerabilities such as CWE-078 (Command Injection) and CWE-200 (Information Exposure) often look benign at the local syntax level—e.g., a system call appears to invoke a legitimate tool, and a logging/return statement seems harmless. 
Only when cross-function data flow and project semantics are considered does it become apparent that untrusted inputs may reach execution sinks or that sensitive data may be propagated to outputs. 
Consequently, both hypothesis construction and hypothesis verification require rich, high-fidelity context (call chains, taint sources/sinks, sanitizers, privilege boundaries). 
In practice, current project context mining is insufficient, which limits VulAgent’s accuracy on these categories: with incomplete context, the agent may either fail to form the correct hypothesis (misses) or lack evidence to confirm/refute it (false alarms).

\textbf{Summary.}  
Overall, while VulAgent achieves state-of-the-art performance at the system level, 
its effectiveness on certain CWE categories remains limited due to insufficient contextual information 
available in the datasets. Combined with the findings of RQ3, this suggests that constructing 
richer and more accurate context is essential for further improving validation performance 
on context-dependent vulnerabilities.

\subsection{RQ5. Effectiveness of Our Hypothesis Validation Process}
\label{sec:rq5}
\begin{table}
\centering
\caption{Effectiveness of our hypothesis validation process compared with baselines.}
\vspace{-1em}
\label{tab:rq5_ablation}
\footnotesize
\setlength{\tabcolsep}{14pt}
\begin{tabular}{lcccccc}
\toprule
Method & P-C$\uparrow$ & P-R$\downarrow$ & VPS$\uparrow$ & FPR$\downarrow$&F1$\uparrow$ & ACC$\uparrow$ \\
\midrule
Direct Evaluation &  36.14& 12.05 &24.10  & 49.40 &65.95 &62.05 \\
VulTrial (Reflection)          & 25.30   & 7.23 & 18.07 & 39.76 & 58.54&59.04 \\
VulAgent (Ours)                & 24.10 &\textbf{1.20} & 22.98 & \textbf{20.48} &52.94 &61.45 \\
\bottomrule
\end{tabular}
\vspace{-1em}
\end{table}

\textbf{Ablation setup.}
We evaluate on SVEN using an \emph{oracle-CWE} setting to isolate the validation stage: each method receives the ground-truth CWE label (following Li et al.~\cite{li2025everything}) along with the same retrieved code context. Thus, performance differences stem solely from the \emph{validation process}.

\begin{itemize}[leftmargin=*]
\item \textbf{Direct Evaluation (CoT).}
Given the oracle CWE label + context, the model makes a single-step vulnerability judgment (no iterative feedback).

\item \textbf{VulTrial (Reflection).}
We replace the original upstream agent outputs with the oracle CWE label, then run the reflection/debate stage to revise the decision.

\item \textbf{VulAgent (Ours).}
We substitute the AggregatorAgent report with the oracle CWE label and execute our validation chain.
\end{itemize}

All three settings use identical code context and scoring protocols; only the validation mechanisms differ.

Table~\ref{tab:rq5_ablation} evaluates three validation strategies under an \emph{oracle-CWE} setting (i.e., detection is bypassed and each method receives the ground-truth CWE). This isolates the validation stage itself.

\textbf{Overall trend.} Once detection noise is removed, all methods see higher accuracy, confirming that misclassification during detection was a confounding factor. However, without a dedicated false-positive control, both one-shot and reflection baselines still produce many spurious alarms.

\textbf{Direct Evaluation.} The one-shot CoT style “Direct Evaluation” achieves the highest P-C (36.14) and the highest F1 (65.95), but suffers from a high FPR of 49.40\%, indicating a strong tendency to over-predict vulnerabilities even with oracle labels.

\textbf{VulTrial (Reflection).} The reflection/debate mechanism reduces FPR to 39.76\% (a 9.6-point drop vs. Direct) and improves VPS compared to naive one-shot prompting, but still leaves a sizable false-positive burden (ACC = 59.04).

\textbf{VulAgent (Ours).} Our hypothesis-validation chain yields the lowest FPR of 20.48\% (a 28.9-point drop vs. Direct and 19.3-point drop vs. Reflection) and the best FPR (20.48\%) with the lowest P-R (1.20), showing stronger pair-wise discrimination and fewer reversed predictions. 
Importantly, this comes with comparable accuracy to Direct (61.45\% vs. 62.05\%, only $-$0.60) and higher accuracy than Reflection (+2.50 pts), demonstrating that our process curbs false positives without materially sacrificing correctness.

 Even when every method is given the same oracle CWE labels, baselines remain prone to false alarms due to the lack of a targeted validation mechanism. By contrast, VulAgent’s hypothesis construction + assumption pruning + protection checking explicitly operationalize context use and deliver a substantial reduction in FPR while preserving accuracy—demonstrating the necessity and effectiveness of the proposed validation process.

\section{Case Study}

To further illustrate the advantage of our approach in reducing false positives, we present a case study on the function \texttt{search\_make\_new} from the \texttt{libevent} project (commit \texttt{ec65c42}, CWE-125). The ground-truth label of this function is \textbf{no-vulnerability}, i.e., no real vulnerability exists. 
We presented the results of the baseline method and the process of hypothesis verification.

\begin{lstlisting}[language=C, 
    caption={Function under analysis: search\_make\_new in evdns.c}, 
    basicstyle=\ttfamily\scriptsize, breaklines=true]
search_make_new(const struct search_state *const state, 
                int n, const char *const base_name) {
    const size_t base_len = strlen(base_name);
    char need_to_append_dot;
    struct search_domain *dom;

    if (!base_len) return NULL;
    need_to_append_dot = base_name[base_len - 1] == '.' ? 0 : 1;

    for (dom = state->head; dom; dom = dom->next) {
        if (!n--) {
            const u8 *const postfix = ((u8 *) dom) + \sizeof(struct search_domain);
            const int postfix_len = dom->len;
            char *const newname = (char *) mm_malloc(
                 base_len + need_to_append_dot + postfix_len + 1);
            if (!newname) return NULL;
            memcpy(newname, base_name, base_len);
            if (need_to_append_dot) newname[base_len] = '.';
            memcpy(newname + base_len + need_to_append_dot, postfix, postfix_len);
            newname[base_len + need_to_append_dot + postfix_len] = 0;
            return newname;
        }
    }

    EVUTIL_ASSERT(0);
    return NULL;
}
\end{lstlisting}

We analyze a representative example in the search\_make\_new function from evdns.c, comparing the baseline single-agent system (VulTrial) with our multi-agent system (VulAgent).

\paragraph{Baseline Result (VulTrial).}
VulTrial reported five distinct vulnerabilities in this function:

\begin{itemize}
\item \textbf{Heap buffer overflow}: It flagged a potential overflow in the line \texttt{memcpy(newname + base\_len + need\_to\_append\_dot, postfix, postfix\_len)}, reasoning that the buffer size might be insufficient if \texttt{postfix\_len} is attacker-controlled.
\item \textbf{Integer overflow}: It assumed that the allocation size expression \texttt{base\_len + need\_to\_append\_dot + postfix\_len + 1} could wrap around and allocate a buffer smaller than required.
\item \textbf{Null dereference}: It flagged possible use of a NULL pointer if \texttt{mm\_malloc} failed to allocate memory for \texttt{newname}.
\item \textbf{Loop-bound failure}: It warned that a negative \texttt{n} value might cause unintended loop behavior.
\item \textbf{Denial-of-service via assertion}: It treated \texttt{EVUTIL\_ASSERT(0)} as attacker-triggerable, potentially causing program termination.
\end{itemize}

However, these findings were based on over-approximated assumptions. The reviewer confirmed all five were false positives:

\begin{itemize}
\item The allocation size is computed with all necessary components, including \texttt{postfix\_len}, ensuring the buffer is large enough for the write.
\item The return value of \texttt{mm\_malloc} is explicitly null-checked before use.
\item The assertion is unreachable under normal conditions and serves as a developer safeguard.
\item The loop variable \texttt{n} is bounded, and the internal list traversal logic ensures correctness.
\item Crucially, \texttt{state->head} and \texttt{dom->len} originate from internal, trusted allocations and cannot be attacker-controlled under any feasible path.
\end{itemize}

These examples illustrate that VulTrial lacks the ability to distinguish between plausible risks and actually exploitable paths.

\paragraph{Our Method (VulAgent).}
VulAgent analyzed the same function using a pipeline of specialized agents (including memory layout, error handling, and symbolic execution). While initial agents surfaced similar risks (e.g., out-of-bounds writes, integer overflows), the FinalValidatorAgent reviewed each trigger path by checking attacker controllability and structural hypotheses. For instance:

\begin{itemize}
\item It traced the allocation and confirmed that \texttt{postfix\_len} is fully included in the buffer size, ensuring safe \texttt{memcpy}.
\item It validated that the null dereference risk is neutralized by an explicit \texttt{if (!newname) return NULL;} check.
\item It rejected the assumption that internal pointers (like \texttt{state->head}) are attacker-controlled, citing structural invariants.
\item It classified the \texttt{EVUTIL\_ASSERT(0)} path as unreachable under attacker influence, confirming it as non-exploitable.
\end{itemize}

All paths were ultimately deemed non-exploitable, and the final output matched the ground-truth label: \textbf{no vulnerability}.

This case reveals a fundamental flaw in LLM vulnerability detectors: they often confuse theoretical flaws with the actual attack surface, thereby resulting in a high rate of false positives.
The current method cannot enable large language models to exclude those attacks that cannot be triggered based on the specific context in engineering projects.
By contrast, VulAgent’s layered reasoning—from symbolic path exploration to assumption validation—enables it to filter out spurious reports effectively. This not only improves precision but also significantly reduces the manual burden on security reviewers.
\section{Discussion}

\textbf{Threats.}
First, our evaluation relies on public datasets (PrimeVul, SVEN) and their pair-wise labels; any label noise or imperfect fix mapping can bias results.
Second, the quality of program context (CFG/DFG/call graph) extracted by static tools (e.g., Joern) may be incomplete due to build options, macros, or missing dependencies, which can affect hypothesis validation.

\textbf{Limitations.}
Our hypothesis–validation paradigm depends on sufficient and relevant context; when inter-procedural or project-level information is missing or noisy, the validator may become conservative and hide true positives.
The multi-stage pipeline introduces extra calls and latency; although we control cost by using non-reasoning models and compact JSON outputs, overhead is non-zero.
Coverage is uneven across CWE families: memory/NULL/authorization patterns are well handled, whereas configuration-, crypto-, or deployment-driven issues are less directly supported.

\section{Related work}
Early systems largely relied on handcrafted rules and pattern templates (e.g., Checkmarx, Flawfinder) to flag suspicious code idioms~\cite{Checkmarx,Flawfinder}. While practical for known patterns, these heuristics demand substantial expert effort and often fail to generalize across projects or coding styles. Moreover, rule-triggering syntactic motifs frequently recur in benign code, which inflates both false positives and false negatives~\cite{yamaguchi2015pattern,yamaguchi2017pattern,li2021sysevr}.

To reduce manual effort, recent work learns vulnerability signals directly from code~\cite{dam2017automatic,russell2018automated}. Broadly, methods fall into two families: sequence encoders and structure-aware models.
\textit{Sequence (token) models} encode code as token streams and learn discriminative features from labeled examples~\cite{russell2018automated,li2018vuldeepecker,cheng2021deepwukong}. Typical designs include CNN/RNN variants and BiLSTM encoders over API-centered “code gadgets”~\cite{schuster1997bidirectional,li2018vuldeepecker}. Their main limitation is a weak grasp of long-range control/data dependencies.
\textit{Graph/structure models} represent programs with AST/CFG/DFG (or related views) and apply GNN-style encoders~\cite{li2021vulnerability,chakraborty2021deep,zheng2021vu1spg,nguyen2022regvd,wu2022vulcnn,cao2022mvd}.

Approaches that leverage pre-trained code models are increasingly adopted for vulnerability detection~\cite{feng2020codebert,guo2020graphcodebert,guo2022unixcoder,niu2022spt,wang2024code,wang2024m2cvd}.
Representative lines of work include: CodeBERT for joint NL–PL pretraining~\cite{feng2020codebert}, GraphCodeBERT which injects data-flow structure via graph-guided masking/attention~\cite{guo2020graphcodebert}, and UniXcoder that unifies code understanding and generation across modalities~\cite{guo2022unixcoder}. Structure-aware pretraining such as SPT further leverages syntactic/semantic signals to strengthen code representations~\cite{niu2022spt}.
Building on these encoders, recent detectors fine-tune for function- or line-level localization, e.g., LineVul/LineVD and Transformer-based variants that adapt PTMs to pinpoint vulnerable lines~\cite{fu2022linevul,thapa2022transformer,hin2022linevd}. 


\paragraph{Single-agent.}
A growing body of work examines whether careful prompting and chain-of-thought (CoT) can turn general LLMs into effective vulnerability detectors~\cite{ding2025vulnerability,nong2024chain,steenhoek2024comprehensive}.
Tamberg and Bahsi benchmark CoT, tree-of-thought, and self-consistency, finding CoT with GPT-4 strongest among prompting variants and competitive with static analysis~\cite{tamberg2025harnessing}. 
Zhou et al. report GPT-4 with prompting outperforming fine-tuned CodeBERT~\cite{zhou2024large}, while Ullah et al. observe solid performance across commercial LLMs but frequent unfaithful reasoning~\cite{ullah2024llms}.
Subsequent analyses question CoT’s reliability for vulnerability tasks~\cite{nong2024chain,steenhoek2024comprehensive,zhang2024prompt}; on the PrimeVul pair set, GPT-3.5/4 perform poorly, underscoring these limits. 

\paragraph{Multi-agent collaboration.}
To address reasoning brittleness and improve specialization, multi-agent designs coordinate complementary roles~\cite{he2025llm}. EvalSVA organizes agents as security specialists~\cite{wen2024evalsva}, and MuCoLD assigns tester/developer roles for collaborative analysis~\cite{mao2024multi}. 
Related ideas appear in smart-contract auditing: LLM-SmartAudit leverages role-play and collaborative reasoning~\cite{wei2024llm}, iAudit uses detector/reasoner/critic roles~\cite{ma2024combining}, and GPTLens deploys multiple prosecutor agents with a critic~\cite{hu2023large}.
VulTrial introduces a courtroom-inspired framework where role-specific agents collaborate to improve automated vulnerability detection~\cite{widyasari2025let}. This method further reduces the false alarm rate by introducing a third party.

\section{Conclusion}
In this paper, we presented \textit{VulAgent}, a multi-agent framework that 
systematically detects and validates software vulnerabilities with the support of 
hypothesis-driven reasoning. 
%
%
This design effectively reduces false positives while retaining the ability 
to capture subtle and context-dependent vulnerabilities.

Comprehensive experiments on two large-scale benchmarks (PrimeVul and SVEN) 
demonstrated the superiority of \textit{VulAgent} over state-of-the-art baselines. 
\textit{VulAgent} achieves balanced precision and recall, 
generalizes across different LLM backends, 
and remains robust under varying amounts of contextual information. 
Ablation studies further confirmed that the proposed 
hypothesis validation mechanism is crucial for improving reliability, 
particularly in mitigating speculative or unreasonable hypotheses. 
%

In the future, we intend to further enhance our work in the direction of context mining. 
Context-based retrieval with more hypothetical conditions may provide more effective and targeted context information for VulAgent.

\begin{acks}
To Robert, for the bagels and explaining CMYK and color spaces.
\end{acks}

\bibliographystyle{ACM-Reference-Format}
\bibliography{sample-base}

\end{document}